\documentclass[12pt]{article}
\usepackage{amsfonts}
\usepackage{amsmath,amssymb}
\newtheorem{thm}{Theorem}

\newtheorem{lem}[thm]{Lemma}
\numberwithin{equation}{section}
\def\Tr{{\rm Tr\,}}

\def\H{{\cal H}}

\def\Det{{\rm Det}}
\def\max{{\rm max \,}}
\def\spec{{\rm spec \,}}

\def\R{\mathbb R}
\def\C{\mathbb C}
\def\Z{\mathbb Z}
\def\N{\mathbb N}
\def\SN{{\cal S}_N}
\def\sgn{{\rm sgn}}

\begin{document}
\title {Random Point Fields for Para-Particles\\
                   of Any Order  }
\author {  Hiroshi Tamura \thanks{tamurah@kenroku.kanazawa-u.ac.jp}\\
          Department of Mathematics, Kanazawa University,\\
          Kanazawa 920-1192, Japan \\
         Keiichi R. Ito \thanks {ito@mpg.setsunan.ac.jp}\\
         Department of Mathematics and Physics, Setsunan University\\
         Neyagawa, Osaka 572-8508, Japan}

\date{\today}

\maketitle
\begin{abstract}
Random point fields which describe gases consisting of para-particles
of any order are given by means of the canonical ensemble approach.
The analysis for the cases of the para-fermion gases are discussed 
in full detail.
\end{abstract}
\section{Introduction}     \label{intro}
Where do the statistics of random point fields come from?
We examine what kind of random point fields follow from 
the para-statistics of particles.

In the previous paper \cite{1}, the boson and/or fermion point fields 
were derived by means of the canonical ensemble approach.
That is, quantum mechanical thermal systems of finite fixed number
of bosons and/or fermions in the bounded box in $\R^d$ were considered.
By taking the thermodynamic limit to the position distribution of 
constituents, random point fields for boson and/or fermion gases of 
positive finite densities and temperatures on $\R^d$ were obtained. 
There, the method was applied to construct the random point fields which 
describe gases consist of para-bosons (resp. para-fermions) of order 2. 
In the recent proceeding article \cite{3}, the argument for para-particle 
gases of order 3 is developed.

In this paper, we pursue the project to the general case: 
we apply the method to statistical mechanics of gases which consist 
of para-particles of any order $p \in \mathbb N$.
We will see that the random point fields obtained in this way are those of 
$\alpha=\pm 1/p $ given in \cite{ST03}.
Our main result in this paper is \\

\smallskip

\noindent
{\bf Theorem} {\sl The random point field for gas of para-fermions 
(resp. para-bosons of low density) of order $p$ is equal in law to the convolution of 
$p$ independent copies of the usual fermion (resp. boson) point field 
for any $p \in \mathbb N$. }

\smallskip

We use the representation theory of the symmetric group
(cf. e.g. \cite{JK,Sa91,Si96}). Its basic facts are reviewed 
briefly in Sect.2, along the line on which the quantum theory 
of para-particles are formulated. 
We state our main results in Sect.3.
Sections 4 and 5 devoted to the full detail of the discussions
on the thermodynamic limits for para-fermions and a few remarks on 
those for para-bosons, respectively.  
Some discussions are given in Sect.6.

\section{Brief review on Representation of the symmetric group}
      \label{symm_group}

We say that 
$ (\lambda_1, \lambda_2, \cdots, \lambda_n) \in {\mathbb N}^n $
is a Young frame of length $n$ for the symmetric group ${\cal S}_N$ if
\[
   \sum_{j=1}^n\lambda_j =N,  \quad
    \lambda_1 \geqslant \lambda_2 \geqslant  \cdots \geqslant \lambda_n
>0.
\]
We associate the Young frame $ (\lambda_1, \lambda_2, \cdots, \lambda_n)$ 
with the diagram of $\lambda_1$-boxes in the first row,
$\lambda_2$-boxes in the second row,..., and $\lambda_n$-boxes 
in the $n$-th row. A Young tableau on a Young frame is a bijection 
from the numbers $1, 2, \cdots, N$ to the $N$ boxes of the frame.

Let $ M_p^N $ be the set of all the Young frames for ${\cal S}_N$ which
have lengths less than or equal to $p$.
For each frame in $ M_p^N $, let us choose one tableau from those on the
frame. The choices are arbitrary but fixed.
${\cal T}_p^N $ denotes the set of all tableaux chosen in this way.
The row stabilizer of a tableau $T$ is denote by ${\cal R}(T)$ , i.e., 
the subgroup of ${\cal S}_N$ consists of those elements that keep all 
rows of $T$ invariant, and ${\cal C}(T)$ the column stabilizer 
whose elements preserve all columns of $T$.

Let us introduce the three elements
\[
   a(T)=\frac{1}{\#{\cal R}(T)}\sum_{\sigma \in {\cal R}(T)}\sigma,
\qquad
   b(T)=\frac{1}{\#{\cal C}(T)}\sum_{\sigma \in {\cal C}(T)}
   \sgn(\sigma)\sigma
\]
and
\[
   e(T)= \frac{d_T}{N !}\sum_{\sigma \in {\cal R}(T)}
   \sum_{\tau \in {\cal C}(T)}\sgn(\tau)\sigma\tau
        = c_Ta(T)b(T)
\]
of the group algebra ${\mathbb C}[{\cal S}_N]$ for each $T \in {\cal T}_p^N$,
where $d_T$ is the dimension of the irreducible representation of
${\cal S}_N$ corresponding to $T$ and 
$ c_T = d_T\times$ $\#{\cal R}(T)\#{\cal C}(T)/N ! $.
As is known,  
\begin{equation}
a(T_1)\sigma b(T_2)=b(T_2)\sigma a(T_1) = 0
\label{asb}
\end{equation}
hold for any $\sigma \in\SN$ if $ T_2 \multimap T_1 $. 
The relations
\begin{equation}
      a(T)^2 = a(T), \quad b(T)^2 =b(T), \quad e(T)^2 =e(T), \quad 
      e(T_1)e(T_2)=0 \quad ( T_1 \neq T_2 )
\label{abe}
\end{equation}
also hold for $ T, T_1, T_2 \in {\cal T}_p^N$.
For later use, let us introduce
\begin{equation}
       d(T) = e(T)a(T) = c_T a(T)b(T)a(T)
\label{defd}
\end{equation}
for  $ T \in {\cal T}_p^N $.
They satisfy
\begin{equation}
      d(T)^2 = d(T), \quad 
      d(T_1)d(T_2)=0 \quad ( T_1 \neq T_2 )
\label{ddd}
\end{equation}
as is shown readily from (\ref{abe}) and (\ref{asb}).
The inner product $\langle \, \cdot\, , \, \cdot \, \rangle$ of $\C[\SN]$ is defined by
\[
    \langle \sigma, \tau \rangle = \delta_{\sigma\tau} \quad \mbox{ for } \;
\sigma, \tau
\in\SN
\]
and the sesqui-linearity.

The left representation $L$ and the right representation $R$ of $\SN$ 
on $\C[\SN]$ are defined by
\[
    L(\sigma)g = L(\sigma)\sum_{\tau\in\SN}g(\tau)\tau
             =\sum_{\tau\in\SN}g(\tau)\sigma\tau
             = \sum_{\tau\in\SN}g(\sigma^{-1}\tau)\tau
\]
and
\[
    R(\sigma)g = R(\sigma)\sum_{\tau\in\SN}g(\tau)\tau
             =\sum_{\tau\in\SN}g(\tau)\tau\sigma^{-1}
             = \sum_{\tau\in\SN}g(\tau\sigma)\tau,
\]
respectively. Here and hereafter we identify $g: \SN \to \C$  with 
$\sum_{\tau\in\SN}g(\tau)\tau \in\C[\SN]$.
They are extended to the representation of $\C[\SN]$ on $\C[\SN]$ as
\[
   L(f)g = fg =\sum_{\sigma,\tau}f(\sigma)g(\tau)\sigma\tau
         =\sum_{\sigma}\big(\sum_{\tau}f(\sigma\tau^{-1})g(\tau)\big)\sigma
\]
and
\[
   R(f)g = g\hat f
         =\sum_{\sigma,\tau}g(\sigma)f(\tau)\sigma\tau^{-1}
         =\sum_{\sigma}\big(\sum_{\tau}g(\sigma\tau)f(\tau)\big)\sigma,
\]
where $\hat f = \sum_{\tau}\hat f(\tau)\tau =
        \sum_{\tau}f(\tau^{-1})\tau
         = \sum_{\tau}f(\tau)\tau^{-1}$.

The character of the irreducible representation of $\SN$ corresponding
to the tableau $T \in {\cal T}_p^N$ is obtained by
\[
    \chi_{T}(\sigma)=\sum_{\tau\in\SN}\langle \tau, L(\sigma) R(e(T))\tau \rangle
    = \sum_{\tau\in\SN}\langle\tau, \sigma \tau \widehat{e(T)}\rangle.
\]
We introduce a tentative notation as in \cite{1}
\begin{equation}
     \chi_{g}(\sigma) \equiv \sum_{\tau\in\SN}\langle\tau, L(\sigma)R(g)\tau\rangle
    = \sum_{\tau,\gamma\in\SN}\langle\tau, \sigma \tau \gamma^{-1}\rangle g(\gamma)
    = \sum_{\tau\in\SN}g(\tau^{-1}\sigma\tau)
\label{chig}
\end{equation}
for $ g=\sum_{\tau}g(\tau)\tau\in\C[\SN]$.
Then $\chi_T = \chi_{e(T)}$ holds.

We consider representations of $\SN$ on Hilbert spaces.
Let $\H_L $ be a certain $L^2$ space which will be specified in the 
next section and $\otimes^N\H_L$ its $N$-fold Hilbert space tensor
product. Let $U$ be the representation of $\SN$ on $\otimes^N\H_L$
defined by
\[
    U(\sigma) \varphi_1\otimes \cdots \otimes \varphi_N =
    \varphi_{\sigma^{-1}(1)}\otimes \cdots \otimes\varphi_{\sigma^{-1}(N)}
     \qquad \mbox{for } \; \varphi_1, \cdots, \varphi_N \in\H_L,
\]
or equivalently by
\[
    (U(\sigma) f)(x_1, \cdots, x_N ) =
    f(x_{\sigma(1)}, \cdots,x_{\sigma(N)})
    \qquad \mbox{ for } \; f \in \otimes^N\H_L.
\]
Obviously, $U$ is unitary: 
   $ U(\sigma)^* = U(\sigma^{-1})=U(\sigma)^{-1}$.
We extend $U$ for $\C[\SN]$ by linearity.
Then $U(a(T))$ is an orthogonal projection because 
$ U(a(T))^* = U(\widehat{a(T)}) = U(a(T)) $ and (\ref{abe}).
So are $U(b(T))$'s, $U(d(T))$'s and 
\begin{equation}
P_{pB} = \sum_{T\in{\cal T}_p^N}U(d(T)). 
\end{equation}
$P_{pB}$ is the projection operator to the subspace for para-bosons 
of order $p$. Note that Ran$\,U(d(T)) = \,$Ran$\,U(e(T))$ because
$d(T)e(T) = e(T)$ and $e(T)d(T)= d(T)$. 

For  para-fermions, we consider the transposed tableau $T'$ of 
$T\in{\cal T}_{p}^{N}$ by exchanging the rows and the columns of 
the Young tableau $T$. 
The transpose $\lambda'$ of frame $\lambda$ is defined in the same way. 
Then $T'$ lives in $\lambda'$ if $T$ lives in $\lambda$. 
 Clearly 
\begin{equation}
{\cal C}(T')={\cal R}(T), \quad {\cal R}(T')={\cal C}(T)
\end{equation}
and we define 
\begin{equation}
P_{pF}=\sum_{T\in{\cal T}_{p}^{N}}U(d(T'))
\end{equation}
which is the projection operator to the subspace for para-fermions
of order $p$.

\section{Para-statistics and Random Point Fields}
\subsection{Para-fermions of Order $p \in \N$} 
\quad We first consider the quantum system of $N$ para-fermions of order $p$
in the box $\Lambda_L = [-L/2, L/2]^d \subset \R^d$. 
We refer the literatures \cite{MG64,HT69,ST70} for quantum mechanics
of para-particles. (See also \cite{OK69}.)
These literatures indicate that the state space of 
our system is given by $\H_{L,N}^{pF} = P_{pF}\otimes^N\!\H_L$,
where $ \H_L = L^2(\Lambda_L) $ with Lebesgue measure is the state
space of one particle system in $\Lambda_L$. 
We need the heat operator $G_L =e^{\beta\triangle_L}$ in $\Lambda_L$,
where $\triangle_L$ is the Laplacian in $\Lambda_L$ with periodic
boundary conditions at $\partial \Lambda_{L}$. Then 
\begin{eqnarray*}
{\rm spec }\, G_{L}&=&\{\exp[-\beta\sum_{j=1}^d q_{j}^{2}];\, 
                   q_{j}\in 2\pi {\mathbb Z}/L\} ,\\
\frac{1}{L^{d}} \Tr G_{L}
            &=& \frac{1}{L^{d}}
            \sum_{n\in {\mathbb Z}^{d}}
            \exp\left(-\beta(2\pi|n|/L)^{2}\right). 
\end{eqnarray*}

It is obvious that there is a CONS of $\H_{L,N}^{pF}$ which 
consists of the vectors of the form 
$U(d(T'))\varphi_{k_1}^{(L)}\otimes \cdots \otimes \varphi_{k_N}^{(L)}$, 
which are the eigenfunctions of $\otimes^NG_L$.
Then, we define a point field of $N$ free para-fermions of order $p$ 
in the box as in section 2 of \cite{1} and its generating functional 
is given by
\[
     E_{L, N}^{pF}\big[e^{-\langle f, \xi\rangle}\big] =
   \frac{\sum_{T\in {\cal T}_p^N}\Tr_{\otimes^N\H_L}
            \big[(\otimes^N\tilde G) U(d(T'))\big]}
        {\sum_{T\in{\cal T}_p^N}\Tr_{\otimes^N\H_L}
             \big[(\otimes^N G) U(d(T'))\big]},
\]
where $f$ is a nonnegative continuous function on $\Lambda_L$ and 
$\tilde G_L = G_L^{1/2}e^{-f}G_L^{1/2}$. We first prove: 
\begin{lem}
  \begin{eqnarray}
  E_{L, N}^{pF}\big[e^{-\langle f, \xi\rangle}\big]
    &=&
   \frac{\sum_{T\in{\cal T}_p^N}\sum_{\sigma\in\SN}\chi_{T'}(\sigma)
         \Tr_{\otimes^N\H_L}[(\otimes^N\tilde G_L) U(\sigma)]}
                {\sum_{T\in{\cal T}_p^N}
         \sum_{\sigma\in\SN}\chi_{T'}(\sigma)
         \Tr_{\otimes^N\H_L}[(\otimes^N G_L) U(\sigma)]}
   \label{fpgfl}    \\
   &=&\frac{\sum_{T\in{\cal T}_p^N}\int_{\Lambda_L^N}
           \det_{T'}\{\tilde G_L(x_i, x_j)\}_{1\leqslant i,j\leqslant N}
                  dx_1 \cdots dx_N}
           {\sum_{T\in{\cal T}_p^N}\int_{\Lambda_L^N}
           \det_{T'}\{ G_L(x_i, x_j)\}_{1\leqslant i,j\leqslant N}
                   dx_1 \cdots dx_N}. 
   \label{fimt}  
   \end{eqnarray}
\label{fl}
\end{lem}
\noindent
{\sl Remark 1 : } The state space $ \H_{L,N}^{pF} = P_{pF}\otimes^N\!\H_L $ is
determined by the choice of the tableaux $T$'s.
The different choices of tableaux give different subspaces of 
$\otimes^N\H_L$.
However, they are unitarily equivalent and the generating functional
given above is not affected by the choice.
In fact, $\chi_{T}(\sigma)$ depends only on
the frame on which the tableau $T$ is defined.

\noindent{\sl Remark 2 : }  det$_{T'}A = \sum_{\sigma\in\SN}
\chi_{T'}(\sigma)\prod_{i=1}^NA_{i\sigma(i)}$ in (\ref{fimt}) is called
immanant.

\noindent {\sl Proof : }  These expressions are derived by the
relations
\[
  \Tr_{\otimes^N\H_L}\big[(\otimes^N G_L) U(d(T'))\big]
   =   \Tr_{\otimes^N\H_L}
             \big[(\otimes^N G_L) U(e(T'))\big]
\]
and
\begin{equation}
 \sum_{\sigma\in\SN}\chi_g(\sigma)
   \Tr_{\otimes^N\H_L}\big[(\otimes^NG_L)U(\sigma)\big]
  =  N!\Tr_{\otimes^N\H_L}\big[(\otimes^NG_L)U(g)\big],
\label{imm}
\end{equation}
with  $g=e(T')$.
These relations can be shown by the use of  (\ref{chig}), the cyclic property 
of the trace and the commutativity of $U(\tau)$ with $\otimes^NG$.  
For details, see \cite{1}. \hfill $\Box$

Now, let us consider the thermodynamic limit
\begin{equation}
       L,\; N \rightarrow \infty, \quad  \rho_{L} \equiv N/L^d \to \rho > 0.
\label{tdl}
\end{equation}
In the following, $f$ is a nonnegative continuous function on $\R^d$ which 
has a compact support, and is fixed through 
the thermodynamic limit $\Lambda_{L}\nearrow\R^{d}$.
We identify the restriction of $f$ to $\Lambda _L$ as $f$ in Lemma \ref{fl}.
We get the limiting random point field on $\R^d$.

\begin{thm}
The finite random point fields for para-fermions of order $p$ defined 
above converge weakly to the point field whose Laplace transform 
is given by
\[
      {\rm E}_{\rho}^{pF}\big[e^{-\langle f, \xi\rangle}\big]
       = \Det \big[1 - \sqrt{1-e^{-f}}r_*G(1 + r_*G)^{-1}
        \sqrt{1-e^{-f}}\big]^p
\]
in the thermodynamic limit (\ref{tdl}), where $r_*\in ( 0, \infty)$ is
determined by
\[
    \frac{\rho}{p} = \int \frac{dp}{(2\pi)^d}\frac{r_*e^{-\beta |p|^2}}
    {1 + r_*e^{-\beta |p|^2}}= (r_*G(1 + r_*G)^{-1})(x,x).
\]
\label{fthm}
\end{thm}

\subsection{Para-bosons of Order $p\in \N $}
\quad We next consider the quantum system of $N$ para-bosons of order $p$
in the box $\Lambda_L = [-L/2, L/2]^d \subset \R^d$.
It is obvious that there is a CONS of $\H_{L,N}^{pB}$ which consists of 
the eigenfunctions for $\otimes^NG_L$ of the form 
$U(d(T))\varphi_{k_1}^{(L)}\otimes \cdots \otimes \varphi_{k_N}^{(L)}$.
Then, we define a point field of $N$ free para-bosons of order $p$ as 
in section 2 of \cite{1} and its generating functional is given by
\[
   E_{L, N}^{pB}\big[e^{-\langle f, \xi\rangle}\big] =
   \frac{\Tr_{\otimes^N\H_L}
            [(\otimes^N\tilde G_L) P_{pB}]}
        {\Tr_{\otimes^N\H_L}
         [(\otimes^N G_L)P_{pB}]},
\]
where $f$ is a nonnegative continuous function on $\Lambda_L$ and 
$\tilde G_L = G_L^{1/2}e^{-f}G_L^{1/2}$. 
Then, we have:
\begin{lem}
  \begin{eqnarray}
  E_{L, N}^{pB}\big[e^{-\langle f, \xi\rangle}\big]
    &=&
   \frac{\sum_{T\in{\cal T}_p^N}\sum_{\sigma\in\SN}\chi_{T}(\sigma)
         \Tr_{\otimes^N\H_L}[(\otimes^N\tilde G_L) U(\sigma)]}
                {\sum_{T\in{\cal T}_p^N}\sum_{\sigma\in\SN}\chi_{T}(\sigma)
         \Tr_{\otimes^N\H_L}[(\otimes^N G_L) U(\sigma)]}
   \label{bpgfl}    \\
   &=&\frac{\sum_{T\in{\cal T}_p^N}\int_{\Lambda_L^N}
           \det_{T}\{\tilde G_L(x_i, x_j)\}dx_1 \cdots dx_N}
           {\sum_{T\in{\cal T}_p^N}\int_{\Lambda_L^N}
           \det_{T}\{ G_L(x_i, x_j)\}dx_1 \cdots dx_N}. 
   \label{bimt}  
   \end{eqnarray}
\label{bl}
\end{lem}

\medskip
We again consider the thermodynamic limit (\ref{tdl}).
We get the limiting random point field on $\R^d$ for the low density
region: 
\begin{thm}
The finite random point fields for para-bosons of order $p$ 
defined above converge weakly to the random point field whose Laplace
transform is given by
\[
      {\rm E}_{\rho}^{pB}\big[e^{-\langle f, \xi\rangle}\big]
       = \Det \big[1 + \sqrt{1-e^{-f}}r_*G(1 - r_*G)^{-1}
       \sqrt{1-e^{-f}}\big]^{-p}
\]
in the thermodynamic limit, where $r_* \in (0, 1)$ is determined by
\[
     \frac{\rho}{p} = \int \frac{dp}{(2\pi)^d}\frac{r_*e^{-\beta |p|^2}}
    {1 - r_*e^{-\beta |p|^2}} = (r_*G(1 - r_*G)^{-1})(x,x),
\]
if
\[
  \frac{\rho}{p} < \rho_c \equiv \int_{\R^d} 
         \frac{dp}{(2\pi)^d}\frac{e^{-\beta |p|^2}}
         {1 - e^{-\beta |p|^2}}.
\]
    \label{bthm}
\end{thm}

\noindent {\sl Remark : } The high density region 
$\rho \geqslant p\rho_c$ 
is related to the Bose-Einstein condensation.
We need a different analysis for the region.
See \cite{2} for the case of $ p= 1$ and $2$. 

\section{Proof of Theorem \ref{fthm} }
                    \label{parafermi}
It is enough to show the convergence of the generating functionals.
In the rest of this paper, we use the results in \cite{1} frequently.
We refer them as, e.g., Lemma I.3.2 for Lemma 3.2 of \cite{1}.
Although those in \cite{1} are results for $p=1$, their arguments hold 
for general $p\in \N$ with obvious changes.
Let $\psi_{T}$ be the character of the induced representation 
Ind$_{{\cal R}(T)}^{\SN}[{\bf 1}] $, where {\bf 1} is the 
one dimensional representation ${\cal R}(T) \ni \sigma \to 1$, i.e.,
\[
         \psi_{T}(\sigma) =\sum_{\tau\in\SN} \langle \tau, L(\sigma)
        R(a(T))\tau\rangle = \chi_{a(T)}(\sigma).
\]
Since the characters $\chi_T$ and $\psi_T$ depend only on the 
frame on which the tableau $T$ lives, not on $T$ itself, we use the notation 
$\chi_{\lambda}$ and $\psi_{\lambda}$ ( $\lambda \in M_p^N$ ) instead of
$\chi_T$ and $\psi_T$, respectively. 

Let $\delta$ be the frame $(p-1, \cdots, 2, 1,0) \in M_p^N$.
Generalize $\psi_{\mu}$ to those $\mu =(\mu_1, \cdots, \mu_p) \in \Z^p$ 
which satisfies $ \sum_{j=1}^p \mu_j = N $ by
\[
    \psi_{\mu} = 0 \qquad \mbox{for } \; \mu \in \Z^p-{\mathbb Z}_+^p
\]
and
\[
   \psi_{\mu} = \psi_{\pi\mu} \quad \mbox{ for } \quad \mu \in {\mathbb Z}_+^p
   \quad  \mbox{ and } \quad \pi\in{\cal S}_p 
 \quad \mbox{ such that } \quad \pi\mu \in M_p^N,
\]
where $ \Z_+ = \{ 0 \} \cup \N$.
Then the determinantal form \cite{JK} can be written as
\begin{equation}
   \chi_{\lambda}  = \sum_{\pi\in{\cal S}_p}\sgn\pi\,
    \psi_{\lambda+\delta-\pi\delta}.
\label{chipsi} 
\end{equation}

Let us recall the relations
\[
     \chi_{T'}(\sigma) = \sgn\,\sigma\,\chi_{T}(\sigma), \qquad
     \varphi_{T'}(\sigma) = \sgn\,\sigma\,\psi_{T}(\sigma),
\]
where
\[
         \varphi_{T'}(\sigma) = \sum_{\tau} \langle \tau, 
        L(\sigma)R(b({T'}))\tau\rangle  = \chi_{b(T')}(\sigma)
\]
denotes the character of the induced representation 
Ind$_{{\cal C}(T')}^{\SN}[ \, \sgn \, ]$, 
where \,sgn\, is the representation 
${\cal C}(T') = {\cal R}(T) \ni \sigma \mapsto \sgn\,\sigma$. 
Then we have a variant of (\ref{chipsi}):
\begin{equation}
 \chi_{\lambda'}  = \sum_{\pi\in{\cal S}_p}\sgn\pi\,
    \varphi_{\lambda'+\delta'-(\pi\delta)'}.
\label{chiphi}
\end{equation}

Now let us consider the denominator of (\ref{fpgfl}).
Let $ T \in {\cal T}_p^N $ live on $ \mu = (\mu_1, \cdots, \mu_p) \in 
M_p^N$. Thanks to (\ref{imm}) for $ g = b(T')$, we have
\[
     \sum_{\sigma\in\SN}\varphi_{T'}(\sigma)
   \Tr_{\!\otimes^N\H_L}\!\big[(\otimes^NG_L)U(\sigma)\big]
   = N!\Tr_{\!\otimes^N\H_L}\!\big[(\otimes^NG_L)U(b(T'))\big]
\]
\[
     = N!\prod_{j=1}^p\Tr_{\!\otimes^{\mu_j}\H_L}\!\big[(\otimes^{\mu_j}G_L)
        A_{\mu_j}\big],
\]
where $A_{n}=\sum_{\tau \in {\cal S}_n}\sgn (\tau) U(\tau)/n! $ is the
anti-symmetrization operator on $ \otimes^n\H_L$.
In the last step, we have used
\[
     b(T') = \prod_{j=1}^p \sum_{\sigma\in{\cal R}_j}\frac{\sgn\sigma}
         {\#{\cal R}_j}\sigma,
\]
where $ {\cal R}_j $ is the symmetric group of $\mu_j$ numbers which
lie on the $j$-th row of the tableau $T$. Now (\ref{chiphi}) yields 
\begin{eqnarray}
\lefteqn{ \sum_{\sigma \in \SN} \chi_{\lambda'}(\sigma)
         \Tr_{\otimes^N\H_L}\!\big[(\otimes^N G_L) U(\sigma)\big]
        } && \nonumber \\
&=&   \sum_{\pi \in{\cal S}_p}\sgn \,\pi\sum_{\sigma\in\SN}
       \varphi_{\lambda'+\delta'-(\pi\delta)'}(\sigma)
           \Tr_{\otimes^N{\cal H}_L}\!
        \big[(\otimes^N G_L)U(\sigma)\big] \nonumber \\
&=& N!\sum_{\pi \in{\cal S}_p}\sgn\, \pi\prod_{j=1}^p 
        \Tr_{\otimes^{\lambda_j -j+\pi(j)}{\cal H}_L}
        \!\big[(\otimes^{\lambda_j -j+\pi(j)} G_L)A_{\lambda_j -j+\pi(j)}\big].
\end{eqnarray}
Here we understand that
$\Tr_{\otimes^n\H_L}\big((\otimes^nG_L)A_n\big)= 1 $ if 
$n=0$ and $=0$ if $ n <0 $ in the last expression.
Applying the Cauchy integral formula to 
\[
    \Det[1 +zJ] =\sum_{n=0}^{\infty}z^{n}
      \Tr_{\otimes^n\H}[(\otimes^n J) A_n]
\]
where $J$ is a trace class operator,  we obtain that 
\begin{equation}
  \Tr_{\otimes^n\H}[(\otimes^nG_{L}) A_n]
    = \oint_{S_{r}(0)}\frac{dz}{2\pi iz^{n+1}}
    \Det[1+z G_{L}],
\label{gIVJ}
\end{equation}
where $S_{r}(\zeta)=\{z\in\C ; \, |z-\zeta|=r\}$. 
Note that $r>0$ can be chosen arbitrary  and the right hand side 
equals $1$ for $ n=0$ and $0$ for $n<0$. Then we have the 
following expression of the denominator of (\ref{fpgfl}): 
\begin{eqnarray}
     &&\sum_{\lambda\in{\cal M}_p^N} \sum_{\sigma \in \SN} 
\chi_{\lambda'}(\sigma)
         \Tr_{\otimes^N\H_L}[(\otimes^N G_L) U(\sigma)]
\notag\\
     &=& N! \sum_{\lambda\in{\cal M}_p^N} \sum_{\pi \in{\cal S}_p}\sgn 
\pi
       \oint\cdots\oint_{S_r(0)^p}\prod_{j=1}^p
       \frac{\Det(1+z_jG_L)\,dz_j}{2\pi i z_j^{\lambda_j -j+\pi(j)+1}}.
\notag\\
     &=& \!\! N! \!\! \sum_{\lambda\in{\cal M}_p^N} \!\!\oint \cdots 
\oint_{S_r(0)^p}
    \!\!\!\! \frac{\Delta_{p}(z_{1},\cdots,z_{p})
    \big[\prod_{j=1}^p \Det(1+z_jG_L)dz_{j}\big]}
       {\prod_{j=1}^p 2\pi i z_j^{\lambda_j + p-j+1}},
\label{pfdenom}
\end{eqnarray}
where $\Delta_{p}(z_{1},\cdots,z_{p})$ is the 
Vandermondian given by
\begin{equation}
   \Delta_{p}(z_{1},\cdots,z_{p}) \equiv 
      \prod_{1 \leqslant i < j \leqslant p}(z_i-z_j)
=\left| \begin{array}{cccc}
         z_{1}^{p-1}   & z_{2}^{p-1}  & \cdots   & z_{p}^{p-1} \\
            \vdots     & \vdots       & \ddots   & \vdots      \\
             z_1       & z_{2}        & \cdots   & z_{p}       \\
              1        & 1            & \cdots   & 1 
       \end{array}
  \right | .
\end{equation}
In the following, we simply write 
$\Delta_{p}(\{z\})$ for $\Delta(z_{1},\cdots,z_{p})$ 
when there is no danger of confusion. 

To make the thermodynamic limit procedure explicit,
we take a sequence $\{L_N\}_{N\in \N}$ which satisfies
$ N/L_N^d \to \rho $ as $ N\to \infty$.
In the following, we set $ r = r_N \in [0, \infty)$ to be the unique 
solution of 
\begin{equation}
   \Tr \frac{rG_{L_N}}{1+rG_{L_N}} = k
\label{r_N}
\end{equation}
where
\begin{equation}
 k=\Big\lfloor \frac{N}{p}+\frac{p-1}{2} \Big\rfloor
\label{kNp}
\end{equation}
is the averaged length of the rows in the Young tableau and 
 $\lfloor \, \cdot \, \rfloor$  represents the integer part.

The existence and the uniqueness of the solution follow from the fact
that  the left-hand side of (\ref{r_N}) is a continuous and  monotone
function of $r$.  For details, see Lemma I.3.2 [for $\alpha = -1 $].
We also put
\begin{equation}
  v_N = \Tr \left[\frac{r_N G_{L_N}}{(1+r_N G_{L_N})^2}\right].
\label{v_N}
\end{equation}
We will suppress the $N$ dependence of $v_N$ and $L_N$.
Since $r_N \to r_*$ in the thermodynamic limit, we have 
$k/(2+r_*) \leqslant v_N \leqslant k$ for large enough $N$.
See Lemma I.3.5. [There $r_N$ and $r_*$ are written as $z_N$ and $z_*$ respectively.]

Put 
\begin{equation}
    (\ref{pfdenom}) = \frac{N! \, \Det[1+rG_{L}]^{p}}
         {(\sqrt{2\pi})^p (\sqrt{v})^{1+p(p-1)/2}r^N}J_p.
\end{equation}
Then we have:

\medskip

\begin{lem} \label{lemJ_p}
\begin{equation}
    \lim_{N \to \infty}J_p =  \frac{1}{p!} \int_{\R^p} 
     |\Delta(y_{1},\cdots,y_{p})|
    \, \delta\Big(\sum_{j=1}^p y_{j}\Big)
     \prod_{j=1}^p e^{-y_{j}^{2}/2}dy_{j} >0. 
\end{equation}
\end{lem}

\medskip

\noindent{\sl Proof : }
We set 
\[
  \nu_{j}= \lambda_{j}+p-j-k. 
\]
Then we have
\begin{subequations}
\begin{eqnarray}
   \nu &=& (\nu_{1},\cdots,\nu_{p})\in {\mathbb Z}^{p}, \\
  \sum_{j=1}^{p}\nu_{j} 
            & = & \nu_{0}   \equiv  N+\frac{p(p-1)}{2}-pk \in[0, p), \\
  \nu_1 &>& \nu_2 > \cdots > \nu_p \geqslant -k.
\end{eqnarray}
\label{nu}
\end{subequations}
The parametrization 
\[
 z_{j} = r\exp(i x_{j}/\sqrt{v}) \qquad \quad ( j=1, \cdots, p )
\]
yields
\begin{subequations}
\begin{eqnarray}
&\Det[1+z_{j}G_{L}]
    = {\rm Det}[1+rG_{L}]
        {\rm Det}\big[1+(z_j-r)G_{L}(1+rG_{L})^{-1}
             \big],\\
&z_{j}-r =
   r(e^{ix_{j}/\sqrt{v}}-1)
          =   r\big(i\sin (x_{j}/\sqrt{v})
                       -2\sin^{2}(x_{j}/2\sqrt{v}) \big),\\
&  dz_{j}
       = ir e^{ix_{j}/\sqrt{v}} dx_{j}/{\sqrt{v}} ,\\
& \Delta_{p}(\{ z \})
       = r^{1+2+\cdots+(p-1)} \Delta_{p}(\{e^{ix/\sqrt{v}}\}) 
\end{eqnarray}
\end{subequations}
and  
\begin{eqnarray}
 J_p &=& \sum_{\nu}
               \bigg( \prod_{j=1}^{p}
              \int_{-\pi\sqrt{v}} ^{\pi\sqrt{v}} \frac{dx_{j}}{\sqrt{2\pi}}
               \bigg)
              \Delta_{p}(\{e^{ix/\sqrt{v}}\})
                  (\sqrt{v})^{1-p+p(p-1)/2}
                \nonumber \\
          &&        \times 
          \bigg(\prod_{j=1}^pe^{-{i}(\nu_{j}+k)x_{j}/{\sqrt{v}}}  
            {\rm Det}\Big[1+(e^{ix_{j}/\sqrt{v}}-1)
         \frac{rG_{L}}{1+rG_{L}}\Big]\bigg),
\end{eqnarray}
where the summation on $\nu$ is taken over all $\nu$ satisfying (\ref{nu}).

We consider two regions of
$x\in(-\sqrt{v}\pi,\sqrt{v}\pi]$
\begin{enumerate}
\item small $x$ region:  $|x| \leqslant v^{1/12}$, 
\item large $x$ region:  $|x| > v^{1/12}$. 
\end{enumerate}
In the large field region, we have 
\begin{eqnarray*}
\lefteqn{
 \bigg|\Det\bigg[1+(z-r)\frac{G_{L}}{1+rG_{L}}\bigg]\bigg|^{2}
        } &&  \\
&=&\Det\Big[1-4\sin^{2}\big(\frac{x}{2\sqrt{v}}\big)\frac{rG_{L}}{1+rG_{L}}
                    \Big( 1-\frac{rG_{L}}{1+rG_{L}}\Big)\Big] \\
&\leqslant&
   {\rm Det}\Big[1-\frac{4}{1+r}
        \sin^{2}\big(\frac{x}{2\sqrt{v}}\big)\frac{rG_{L}}{1+rG_{L}}
             \Big] \\
&\leqslant& \exp\Big(-\frac{4}{1+r} \, 
        \sin^{2}\big(\frac{x}{2\sqrt{v}}\big) \Tr \frac{rG_{L}}{1+rG_{L}}\Big)
        \leqslant \exp(-{\rm const}  \, N^{1/6}),
\end{eqnarray*}
using 
$  0 \leqslant G_{L} \leqslant 1 $
and $ v=O(k)=O(N)$ and the boundedness of $r = r_N >0$ uniformly in $N$.

In the small field region, we have the convergent expansion
\begin{eqnarray}
\lefteqn{
       \Det\Big[1+(z-r)\frac{G_{L}}{1+rG_{L}}\Big]
        } && \nonumber \\
  && = \exp\bigg[(e^{ix/\sqrt{v}}-1)k 
             -\frac{1}{2}(e^{ix/\sqrt{v}}-1)^{2}(k-v)
                \nonumber  \\
  &&  \mbox{\hspace*{3cm}}+ \sum_{\ell=3}^{\infty}
             \frac{(-1)^{\ell-1}}{\ell}
              (e^{ix/\sqrt{v}}-1)^{\ell} 
             \Tr \left(\frac{rG_{L}}{1+rG_{L}}\right)^{\ell} \bigg]
                \nonumber \\
 &&= \big( 1+\sum_{\ell=3}^{n-1} c_{\ell}{x}^{\ell}  + R_n(x)\big)
      \exp\Big( \frac{ikx}{\sqrt{v}}- \frac{1}{2}x^{2}\Big),
\label{Remainder}
\end{eqnarray}
where
\begin{eqnarray}
   | c_{\ell} | &\leq& {\rm const }\,  N^{-\ell/6}, \qquad
   ( \, \ell = 3, \cdots, n-1 \, )
\label{BoundForC}
\end{eqnarray}
and
\[
    ||R_n||_{\infty} \equiv \sup_{|x|\leqslant v^{1/12} }|R_n(x)| 
         = O\big(N^{-n/12}\big)
\]
hold.
We put
$$
\sum_{\ell=3}^{n-1} c_{\ell}{x}^{\ell}  = \delta(x).
$$
We choose $n$ in (\ref{Remainder}) so large that 
\begin{equation}
       \sum_{\nu}(\sqrt v)^{1-p+p(p-1)/2}||\Delta_p||_{\infty}||R_n||_{\infty}
         =o(1)
\label{Cond.for.n}
\end{equation}
holds, i.e., $ n> 3(p-1)(p-2) $.

These arguments show that it is enough to consider the contribution
from the small $x$ region, and we have
\begin{eqnarray}
   J_p &=&  \bigg\{
              \sum_{\nu}
               \bigg( \prod_{j=1}^{p}
              \int_{-v^{1/12}} ^{v^{1/12}} \frac{dx_{j}}{\sqrt{2\pi}}
           \, e^{-i\nu_jx_j/\sqrt{v} - x_{j}^{2}/2}(1+\delta(x_j))
              \bigg)        \nonumber   \\
          && \times   {\Delta_{p}(\{e^{ix/\sqrt{v}}\})}
              (\sqrt{v})^{1-p+p(p-1)/2}\bigg\}
                 +o(1),     \nonumber \\
          &=&  \bigg\{
              \sum_{\nu}  \bigg( \prod_{j=1}^{p}
              \Big(1+\delta\Big(i\sqrt v \frac{\partial}{\partial\nu_j}\Big)\Big)
              \int_{-\infty} ^{\infty} \frac{dx_{j}}{\sqrt{2\pi}}
           \, e^{-i\nu_jx_j/\sqrt{v} - x_{j}^{2}/2}  
              \bigg)    \nonumber \\
         &&  \times   {\Delta_{p}(\{e^{ix/\sqrt{v}}\})}
                 (\sqrt{v})^{1-p+p(p-1)/2}\bigg\}
                  +o(1).
\label{fullJ_p}
\end{eqnarray} 

Thanks to the multi-linearity of the determinant $\Delta_{p}$,
we have
\begin{eqnarray}
\lefteqn{ \bigg( \prod_{j=1}^{p}
              \int_{-\infty} ^{\infty} \frac{dx_{j}}{\sqrt{2\pi}}
           \, e^{-i\nu_jx_j/\sqrt{v} - x_{j}^{2}/2}  
              \bigg)
              {\Delta_{p}(\{e^{ix/\sqrt{v}}\})}}&&
\nonumber \\
&=& \det\bigg\{ \int_{-\infty} ^{\infty}
  \frac{dx_{j}}{\sqrt{2\pi}}
     e^{i(l-\nu_j)x_j/\sqrt{v}
              - x_{j}^{2}/2}\bigg\}_{\!\!\!\!\mbox{\tiny
                      \begin{tabular}{l}
                      $p-1  \geqslant \ell \geqslant 0$ \\
                      $1    \leqslant j    \leqslant p$  
                     \end{tabular}
                    }} \nonumber \\
 &=&   \det \Big\{e^{-(\ell-\nu_{j})^{2}/2v} \Big\}_{\!\!\!\!\mbox{\tiny
                      \begin{tabular}{l}
                     $p-1  \geqslant \ell \geqslant 0$ \\
                      $1   \leqslant  j   \leqslant p$  
                     \end{tabular}
                     }}  \nonumber \\
 &=&  \Delta\big(\{e^{\nu/v}\}\big)
                       \exp\!\bigg(\!-\sum_{j=1}^p\frac{\nu_{j}^{2}}{2v} 
       - \sum_{\ell=0}^{p-1}\frac{\ell^{2}}{2v}\bigg).
\label{VanInteg}
\end{eqnarray}
Since $(\sqrt v)^{p(p-1)/2}\Delta_p(\{e^{\nu/v}\})
   = \Delta_p(\{\sqrt v(e^{\nu/v}-1)\})$, 
the summation over all $\nu$ satisfying the condition (\ref{nu}) yields
\[
    \lim_{N\to\infty}\sum_{\nu} (\sqrt{v})^{1-p+p(p-1)/2}\times(\ref{VanInteg}) 
\]
\begin{equation}
      = \int_{y_{1}>\cdots >y_{p}} 
     \delta(\sum_{j=1}^p y_{j})
     \Delta(y_{1},\cdots,y_{p})
     \prod_{j=1}^p e^{-y_{j}^{2}/2}dy_{j}.
\label{mainterm}
\end{equation}
Here we have put $y_j=\nu_j/\sqrt v$ and regarded 
$(\sqrt v)^{1-p}\sum_{\nu}$ as the integral of the suitable step function. 
Then, $(\sqrt v)^{1-p}\sum_{\nu}\to \int dy \,\delta(\sum_jy_j)$
is derived by the use of the dominated convergence theorem.
The limit of the main term of (\ref{fullJ_p}) is given by (\ref{mainterm}),
which is equal to (\ref{lemJ_p}).
We may see that the limit of the remainder vanishes from
(\ref{BoundForC}) and the convergence of 
\[
  \sum_{\nu}  \bigg( \prod_{j=1}^{p}(\sqrt v) ^{a_j} 
          \frac{\partial^{a_j} }{\partial\nu_j^{a_j} }
              \int_{-\infty} ^{\infty} \frac{dx_{j}}{\sqrt{2\pi}}
    \, e^{-i\nu_jx_j/\sqrt{v} - x_{j}^{2}/2} \bigg )
\]
\[
          \times \bigg.  {\Delta_{p}(\{e^{ix/\sqrt{v}}\})}
                 (\sqrt{v})^{1-p+p(p-1)/2} 
\]
\begin{eqnarray}
&=&\sum_{\nu} (\sqrt{v})^{1-p}
          \Big( \prod_{j=1}^{p}(\sqrt v) ^{a_j} 
          \frac{\partial^{a_j} }{\partial\nu_j^{a_j} }
          \Big )
       \Delta_p\big(\{\sqrt v(e^{\nu/v}-1)\}\big)
       \exp\!\Big(\!-\sum_{j=1}^p\frac{\nu_{j}^{2}}{2v} 
       - \sum_{\ell=0}^{p-1}\frac{\ell^{2}}{2v}\Big)\nonumber \\
&\longrightarrow&
     \int_{y_{1}>\cdots >y_{p}} 
     \delta(\sum_{j=1}^p y_{j})
       \Big(\prod_{j=1}^{p}\frac{\partial^{a_j} }{\partial y_j^{a_j} }\Big)
     \Delta_p(y_1, \cdots, y_p)
    e^{-\sum_{j=1}^{p}y_{j}^{2}/2} 
      \Big(\prod_{j=1}^{p}dy_{j}\Big).
\label{Riemann}
\end{eqnarray}
We obtain this convergence by performing the differentiations in the
second  and the third members of (\ref{Riemann}) and applying 
the dominated convergence theorem.
\hfill $\square$

\medskip

The numerator is obtained just in the same way. That is,
we replace
$G_{L}$ by $\tilde{G}_{L}=G_{L}^{1/2}e^{-f}G_{L}^{1/2}$ and 
introduce $\tilde{r}=\tilde{r}_{N}$ and $\tilde{v}=\tilde{v}_{N}$ by
\[
             \Tr \frac{\tilde{r_N}\tilde{G}_{L_N}}
         {1+\tilde{r_N}\tilde{G}_{L_N}}=k,  \qquad
          \Tr \frac{\tilde{r_N}\tilde{G}_{L_N}}
         {(1+\tilde{r_N}\tilde{G}_{L_N})^2}= \tilde v_N.
\]
$G_L-\tilde{G_L}$ is a positive operator of trace class 
such that $\Tr(G-\tilde{G})$
$=O(||1-e^{-f})||_1)$ and 
$\max\{\spec G_{L}\}-\max\{\spec \tilde{G}_{L}\} =  O(L^{-d})$.
Then it follows from the definitions of $r_{N}, \tilde{r}_{N}, v_{N}$
and $\tilde{v}_{N}$ that 
\begin{equation}
     0 \leqslant  \tilde{r}_{N}-r_{N} = O(N^{-1}),\quad
    |\tilde{v}_{N}-v_{N}| = O(1). 
\end{equation}
See Lemma I.3.5 and Lemma I.3.6 for detail.

We define $\tilde J_p$ similarly ($\tilde{v}_{N}$ is used instead 
of $v_{N}$) and we get
\[
      \lim_{N \to \infty}\tilde J_p =  \frac{1}{p!} \int_{\R^p} 
     |\Delta_p(y_{1},\cdots,y_{p})|
    \, \delta\Big(\sum_{j=1}^p y_{j}\Big)
     \prod_{j=1}^p e^{-y_{j}^{2}/2}dy_{j} >0
\]
by the very same argument as in the proof of Lemma \ref{lemJ_p}.

Thus we have (writing $r=r_{N}$, $\tilde{r}=\tilde{r}_{N}$ and $L=L_N$)
\begin{eqnarray}
(\ref{fpgfl})&=&
\bigg(\frac{{\rm Det}[1+\tilde{r}\tilde{G}_{L}]}
           {{\rm Det}[1+rG_{L}]}\bigg)^{p} 
          \Big(\frac{r}{\tilde{r}}\Big)^{N}
          \Big(\frac{v}{\tilde{v}}\Big)
               ^{p(p-1)/4+1/2} 
           \frac{\tilde J_p}{J_p}      \nonumber \\
&=& \bigg(\frac{{\rm Det}[1+r\tilde{G}_{L}]}
       {{\rm Det}[1+rG_{L}]}\bigg)^{p} 
     {\rm Det}
    \Big[1+\frac{(r-\tilde{r})}
           {1+\tilde r \tilde{G}_{L}}\tilde{G}_{L}\Big]^{-p}  \nonumber \\
& & \times 
        \Big(1+\frac{r-\tilde{r}}{\tilde{r}}\Big)^N
        \Big(\frac{v}{\tilde{v}}\Big)^{p(p-1)/4+1/2}\frac{\tilde{J_p}}{J_p}.
\label{flimit}
\end{eqnarray}
Here
\[
       \bigg(\frac{{\rm Det}[1+r\tilde{G}_{L}]}
           {{\rm Det}[1+rG_{L}]}\bigg)^{p} 
 = \Det \Big[1+\frac{r}{1+rG_{L}}(\tilde{G}_{L}-G_{L})
         \Big]^{p}  
\]
\[
         = {\rm Det}\Big[1-\sqrt{1-e^{-f}}
           \frac{rG_{L}}{1+rG_{L}}\sqrt{1-e^{-f}}\Big]^{p}
\]
\[
          \to {\rm Det}\Big[1-\sqrt{1-e^{-f}}
           \frac{r_*G}{1+r_*G}\sqrt{1-e^{-f}}\Big]^{p}
\]
holds. For details, we refer Proposition I.3.9 (and the argument on (c) in the 
proof of Theorem I.3.1).
The remaining factor of (\ref{flimit}) tends to 1 as $N\to \infty$ 
since $v/\tilde{v} \to 1, \tilde J_p/J_p \to 1$ and 
\begin{eqnarray*}
 && N\log\Big(1+\frac{r-\tilde r}{\tilde r} \Big)
           -p\log\Det\Big[1+\frac{(r-\tilde{r})}
           {1+\tilde r \tilde{G}_{L}}\tilde{G}_{L}\Big] \\
   &=&  N\frac{r-\tilde{r}}{\tilde{r}}
  - p \frac{r- \tilde{r}}{\tilde{r}}
   \Tr \frac{\tilde{r}\tilde{G}_{L}}{1+\tilde r\tilde{G}_{L}} +O(N^{-1}) \\
&=&  \frac{r-\tilde{r}}{\tilde{r}}(N-pk) 
     + O(N^{-1}) = O(N^{-1}).
\end{eqnarray*}

Finally, we also get
\[
  \frac{\rho}{p} = \lim_{N\to\infty}
                   \frac{1}{L_N^{d}}
                   \Tr \frac{r_NG_{L_N}}{1+r_NG_{L_N}}
               =  (r_*G(1 + r_*G)^{-1})(x,x)
\]
from (\ref{r_N}), (\ref{kNp}), $N/L_N^d \to \rho$ and Proposition I.3.9.
This completes the proof of Theorem \ref{fthm}. \hfill $\square$

\section{Proof of Theorem \ref{bthm} }
          \label{paraboson}
In the case of para bosons, immanants (permanents) 
$\sum_{\sigma\in S_{N}}\prod_{j=1}^N G_{L}(x_{j},x_{\sigma(j)})$
and
$\sum_{\sigma\in S_{N}}\prod_{j=1}^N \tilde{G}_{L}(x_{j},x_{\sigma(j)})$
appear in the denominator and the numerator, respectively. 
We can represent them by the Fredholm determinants
by making use of Vere-Jones' formula \cite{VJ,ST03}:
\[
    \Tr_{\otimes^n\H}[(\otimes^nG_{L}) S_n]
    = \oint_{S_{r}(0)}\frac{dz}{2\pi iz^{n+1}}
    \Det[1 - z G_{L}]^{-1},
\]
where $S_{n}=\sum_{\tau \in {\cal S}_n}U(\tau)/n! $ 
is the symmetrization operator on $ \otimes^n\H_L$ and $r \in (0,1)$ 
in this case.

Using (\ref{chipsi}), we get the following expression 
of the denominator of (\ref{bpgfl}): 
\begin{eqnarray}
     &&\sum_{\lambda\in{\cal M}_p^N} \sum_{\sigma \in \SN} 
\chi_{\lambda}(\sigma)
         \Tr_{\otimes^N\H_L}[(\otimes^N G_L) U(\sigma)]
\notag\\
     &=& N! \sum_{\lambda\in{\cal M}_p^N} \sum_{\pi \in{\cal S}_p}\sgn 
\pi
       \oint\cdots\oint_{S_r(0)^p}\prod_{j=1}^p
       \frac{dz_j}{2\pi i z_j^{\lambda_j -j+\pi(j)+1}
                    \Det[1-z_jG_L]\,}
\notag\\
     &=& \!\! N! \!\! \sum_{\lambda\in{\cal M}_p^N} \oint \cdots 
\oint_{S_r(0)^p}
    \!\! \frac{\Delta_{p}(z_{1},\cdots,z_{p})\,dz_1 \cdots dz_p}
    {\big(\prod_{j=1}^p 2\pi i z_j^{\lambda_j + p-j+1} \Det[1-z_jG_L]\big)},
\label{pb_denom}
\end{eqnarray}
where $\Delta_{p}(z_{1},\cdots,z_{p})$ is the 
Vandermondian introduced in the previous section. 

We choose a sequence $\{L_N\}_{N\in \N}$ which satisfies
$ N/L_N^d \to \rho $ as $ N\to \infty$.
In this case, $ r_N \in (0, 1)$ denotes the unique 
solution of 
\begin{equation}
   \Tr \frac{rG_{L_N}}{1-rG_{L_N}} = k
\end{equation}
where
\begin{equation}
 k=\lfloor\frac{N}{p}+\frac{p-1}{2} \rfloor
\end{equation}
as in (\ref{kNp}).
We put
\begin{equation}
       v_N = \Tr \Big[\frac{r_NG_{L_N}}{(1-r_NG_{L_N})^2}\Big].
\end{equation}
The remaining parts are almost the same as those in the 
para-fermion case. The reader may complete the proof of 
Theorem \ref{bthm}, following the previous arguments with
the obvious changes. 

\section{Discussion}
         \label{disc}

We have shown that

\smallskip
  
\noindent the generating functional of the gas of para-fermions 
 (resp. para-bosons of low density) of order $p$ is equal to the $p$-th power 
 of the generating functional of fermions (resp. bosons).
 
\smallskip

The random point fields which we have
obtained in this paper are a subset of those in \cite{ST03}, where 
various properties of the point fields are examined.
On the other hand, the authors of \cite{ST03} obtained the point 
fields which do not follow from the representation theory of 
the symmetric groups which we discussed in this paper.
Therefore it is interesting to consider physical interpretations 
of the point fields which do not follow from the 
representation theory of the symmetric groups. See e.g. \cite{Wil}.

\bigskip

\noindent
{\sl Acknowledgements.}
We would like to thank Professors Y.Takahashi and T. Shirai for
useful discussions.
H.T. is grateful to the Grant--in--Aid 
for Science Research No.17654021 from MEXT. 
K.R.I. would like to thank the Grant--in--Aid 
for Science Research (C)15540222 from JSPS.


\end{document}